\documentstyle[aps,prl,twocolumn]{revtex}
\input{epsf}
\begin{document}
\draft
\preprint{\today}
\setcounter{totalnumber}{99}
\renewcommand{\textfraction}{0}
\title{Signatures of electron correlations in the transport properties of
quantum dots.}
\author{Kristian Jauregui\thanks{E.mail: jauregui@physnet.uni-hamburg.de},
Wolfgang H\"ausler, Dietmar Weinmann$^{\mbox{\footnotesize\S}}$ and
Bernhard Kramer}

\address{I. Institut f\"ur Theoretische Physik, Jungiusstr.~9, 20355
Hamburg, F.R.G.\\ $^{\mbox{\footnotesize\S}}$CEN Saclay, SPEC,
F--91191 Gif--sur--Yvette Cedex, France.}
 \vspace{1cm}
\maketitle
\date{}
\thispagestyle{empty}
\begin{abstract}
The transition matrix elements between the correlated $N$ and $N\!+\!1$
electron states of a quantum dot are calculated by numerical
diagonalization. They are the central ingredient for the linear and
non--linear transport properties which we compute using a rate equation. The
experimentally observed variations in the heights of the linear conductance
peaks can be explained. The knowledge of the matrix elements as well as the
stationary populations of the states allows to assign the features observed
in the non--linear transport spectroscopy to certain transition and contains
valuable information about the correlated electron states.

\end{abstract}


\narrowtext

By using modern nanostructure fabrication technology a few electrons can be
confined to very small regions in space \cite{kastner92}. In these so--called
quantum dots or artificial atoms the Coulomb interaction between the
electrons is very important for understanding their quantum mechanical
properties. Weak coupling to external reservoirs via tunnel barriers allows
to observe single electron transport effects like the Coulomb blockade
oscillations in the linear conductance at millikelvin temperatures
\cite{meirav90,johnson92,mceuen93}. In non--linear transport, features  are
observed which are closely related to the excitation spectrum of the
interacting electrons \cite{weis93}.

Transport involves transitions between the many--body eigenstates of the
confined electrons. They are approximated as products of one particle states
within the charging model where the Coulomb interaction is modelled
phenomenologically by the capacity of the quantum dot \cite{grabert91}. This
is not sufficient to explain e.g. the experimentally observed negative
differential conductances \cite{pfaff94}. Especially at low electron
densities, correlations between the electrons are crucial. This was
explicitly shown for a quasi one--dimensional (1D) box \cite{kris93} where
the correlation leads to $N$ pronounced peaks in the charge density
distribution if the mean electron distance $r_{\rm s}$ exceeds the effective
Bohr radius $a_{\rm B}^{*}$ with $a_{\rm B}^{*}\equiv(m_{\rm
e}/m)\varepsilon a_{\rm B}$ ($\varepsilon$ relative dielectric constant, $m$
effective mass). In the present paper we use the same model to investigate
the influence of the spatial properties of the correlated many--electron
states on the linear and non--linear transport properties of an artificial
atom. An investigation in a similar spirit was recently performed for a
harmonic confining potential in 2D \cite{daniela95}. However transport
properties were not calculated and only the transition between $N\!=\!2$ and 3
electron states was studied in this work. We demonstrate here that the
current--voltage characteristics obtained by solving a stationary master
equation exhibits very specific signatures of the electron correlations.
They influence the transition matrix elements and also stationary occupation
probabilities of the states. Some of the ''lines'' observed in non--linear
transport spectra are even enhanced. Such a result cannot be obtained by
considering only the transitions rates.

We will show that apart from the restrictions due to spin selection rules
discussed earlier \cite{dietmar94,dietmar95} transitions are suppressed or
enhanced when taking into account the spatial properties of the
wave--functions. The heights of the peaks in the linear conductance become
non--equal even without taking into account the energy dependence of
tunneling matrix elements. In non--linear transport the excited levels of
the quantum dot which {\it can} be observed are closely related to the {\it
most prominent allowed transitions} and to the {\it highest stationary
occupation probabilities}.

As a model for the quantum dot we consider a quasi--1D square well
\cite{kris93,wolfg93} of length $L=9.5a_{\rm B}^{*}$ and $N \le 4$. The
corresponding mean electron density is close to the one in experiments on
GaAs--AlGaAs heterostructures where the mean distance between the electrons
is about $3 a_{\rm B}^{*}\:$ ($\approx$10nm)
\cite{meirav90,mceuen93,weis93}.

Including the spin degree of freedom $\sigma$, the dot Hamiltonian reads
 \begin{eqnarray} \label{hd}
H_{\rm D} & = & \sum_{n,\sigma}(\epsilon_{n}-e \phi)
c^{\dagger}_{n,\sigma}c^{\phantom{\dagger}}_{n,\sigma} \nonumber\\
& + & \sum_{n_{1}...n_{4} \atop\sigma,\sigma'}
V_{n_{4}n_{3}n_{2}n_{1}}^{\phantom{\dagger}} c_{n_{4}\sigma}^{\dagger}
c_{n_{3}\sigma'}^{\dagger}c_{n_{2}\sigma'}c_{n_{1}\sigma}\; .
 \end{eqnarray}
The electrostatic potential $\phi$ depends on the gate and transport (bias)
voltages which are applied to the system and shifts the energies of the
one--electron levels. $V_{n_{4}n_{3}n_{2}n_{1}}$ is the matrix  element of the
interaction $V(x,x')={e^2}/{\varepsilon\sqrt{(x- x')^2+ \lambda^2}}$. The
cutoff at short distances simulates a small transversal spread of the wave
functions ($\:\lambda\ll L\:$) and provides finite
$V_{n_{4}n_{3}n_{2}n_{1}}$.

In the limit of high tunnel barriers, transport is determined by the
eigenstates of the isolated dot. From the latter we calculate the transition
probabilities between $N$ and ($N\pm1$) electron states. Each energy
eigenstate is simultaneously eigenstate of the total spin $\hat{S}^{2}=
(\hat{\sigma}_{1}+\cdots+\hat{\sigma}_{\rm N})^{2}$ which implies a
($2S\!+\!1$)--fold degeneracy with respect to $\hat{S}_{z}$ in the absence
of a magnetic field.

We include $n\!=\!1,\ldots,M$ one--electron states
$\:\varphi_{n}(x)\chi_{\sigma}\:$ when diagonalizing $H_{\rm D}$. Here,
$\varphi_{n}$ is a spatial one electron function and $\chi_{\sigma}$ a
spinor with $\:\sigma=\downarrow,\uparrow\:$. The Hamiltonian matrix in the
basis of Slater determinants is of the rank of the binomial number
$r=C_{2M}^{N}$
($r<1.5\cdot 10^4$, for $\:M=10,\ldots ,13\:$). The Lanczos method was used
when $\:r>3\cdot 10^{3}\:$.

By its algorithm \cite{lanczos} the Lanczos diagonalization provides only
one eigenvector $|\Psi^{S}\rangle_{\rm{Lanc}}$ for each energy eigenvalue.
The calculation of transition rates is considerably facilitated when using
eigenstates of $\hat{S}_z$. Usually $|\Psi^{S}\rangle_{\rm{Lanc}}$ is a
linear combination of all of the $2S\!+\!1$ vectors in the subspace of
Zeeman levels. In order to recover the eigenvectors of $\hat{S}_z$ we apply
projectors $\hat{P}_{S_{z}}|\Psi^{S}\rangle_{\rm Lanc}\propto
|\Psi^{S,S_{z}}\rangle$ corresponding to a specific $S_{z}$. After
normalization $|\Psi^{S,S_{z}}\rangle\equiv\sum_{\nu=1}^{r}
\mbox{b}_{\nu}^{S,S_{z}}|\nu\rangle$ can be expanded into the Slater
determinants $|\nu\rangle=c_{n_{1}\sigma_{1}}^{\dagger}\cdots
c_{n_{N}\sigma_{N}}^{\dagger}|0\rangle$ of the noninteracting electrons. The
coefficients $\mbox{b}_{\nu}^{S,S_{z}}$ are obtained after diagonalization,
projection and renormalization. By construction we have
$\mbox{b}_{\nu}^{S,S_{z}}=0$ if $\sigma_{1}+\cdots+\sigma_N\neq
S_{z}$. The method would fail in the unlikely case that
$|\Psi^{S}\rangle_{\rm Lanc}$ is accidentally perpendicular (within
the numerical accuracy) to one of the $|\Psi^{S,S_{z}}\rangle$. The
procedure can also be applied to higher dimensional models.

To study transport properties, we use the tunneling Hamiltonian and the
usual rate equation approach \cite{dietmar94,dietmar95}. Then the
dc--current through the quantum dot
 \begin{equation}\label{current}
I\equiv I^{\rm L/R}=(-/+) e\sum_{i,j\, (j\neq i)}P_{i}
\Gamma^{\rm L/R}_{i,j} (N_j - N_i)
 \end{equation}
corresponds to the rate of electron passages through the left or the right
barrier. $I$ is computed from the stationary occupation probabilities $P_j$
which are solutions of the equation $\sum\limits_{j\, (j\neq i)}
(\Gamma_{i,j}P_{j}-\Gamma_{j,i}P_{i})=0$. The transition rates between all
of the many--electron states indexed by $j$ are
$\Gamma_{j,i}=\Gamma_{j,i}^{\rm L}+\Gamma_{j,i}^{\rm R}$.

Time dependent perturbation theory yields
 \begin{eqnarray}\label{matrat}
\Gamma_{j,i}^{\rm L/R} & = & t^{\rm L/R} \left| \sum_{n,\sigma} \langle
\Psi_{j}| c_{n,\sigma}^{\dagger} |\Psi_{i}\rangle \right|^{2} f^{\rm L/R}(E)
\nonumber\\
& + & t^{\rm L/R} \left| \sum_{n,\sigma} \langle
\Psi_{i}|c_{n,\sigma}^{\phantom{\dagger}} |\Psi_{j}\rangle \right|^{2} [1 -
f^{\rm L/R}(-E)]\; ,
 \end{eqnarray}
for transition probabilities between the eigenstates $|\Psi_{i}\rangle$ and
$| \Psi_{j}\rangle$ of $H_{\rm D}$ with $N_{j}=N_{i}+1$ in lowest order in
the tunneling. The indices $i$ and $j$ contain in particular the
electron number $N$, the total spin $S$ and $S_{z}$. $t^{\rm L/R}$ are the
tunneling rates through the left/right barrier. The electron has to provide
the energy difference $E = E_{j} - E_{i}$ when entering or leaving the dot.
The Fermi-Dirac distribution functions $f^{\rm L/R}(E)$ describe the
left/right reservoirs with chemical potentials $\mu^{\rm L/R}$.

The energy spectrum of the $N$ electrons for the densities studied consists
of multiplets \cite{kris93,wolfg93}. The energy differences $\Omega$ between
the latter are considerably larger than the intra--multiplet energy
differences $\Delta$. They are important only for large transport voltages.
For a GaAs--AlGaAs heterostructure of length $L=9.5a_{\rm B}^{*}$ with
$N=4$, $\Omega\approx6.2meV$, and $\Delta\approx62\mu eV$. For small
transport voltages we can restrict ourselves to transitions between states
within the lowest multiplets (table \ref{spectrum}). The total number of
states within the multiplets, including the $S_{z}$--degeneracy, is $2^{N}$.

In the following we discuss the influence of the correlations between the
electrons on the total transition probability
 \begin{equation}\label{overmatel}
M_{j,i}=\frac{1}{2}\;\frac{1}{2S_i+1}\sum_{S_{zi}=-S_i}^{S_i}
\sum_{S_{zj}=-S_j}^{S_j} \left|\sum_{n,\sigma} \langle \Psi_{j}|
c_{n,\sigma}^{\dagger} |\Psi_{i}\rangle\right|^{2}\; ,
 \end{equation}
where the spins $S_{zi}$ and $S_{zj}$ refer to the states $|\Psi_{i}\rangle$
and $\langle\Psi_{j}|$, respectively. The matrix elements $\langle \Psi_{j}|
\,c_{n,\sigma}^{\dagger} |\Psi_{i}\rangle$ imply first of all a spin selection
rule, namely that each added or removed electron can change both the total
spin $S$ and the magnetic quantum number $S_{z}$ only by $\pm1/2$. In
\cite{dietmar94,dietmar95} these selection rules were included via the
Clebsch--Gordan coefficients. They describe the combination of the initial
spin state ($S_i,S_{zi}$) with an electron ($1/2,\pm1/2$) to form the final
state ($S_j,S_{zj}$) and yield
 \begin{equation}\label{clebgor}
C_{j,i}=\frac{S_i+1}{2S_i+1}\delta_{S_j,S_i+1/2} +
\frac{S_i}{2S_i+1}\delta_{S_j,S_i-1/2}
 \end{equation}
after summation over $S_{zi}$ and $S_{zj}$. This approach ignores the
spatial degrees of freedom which make the $M_{j,i}$ considerably different
from the $C_{j,i}$.

In tables \ref{table23} and \ref{table34} the numerically calculated
$M_{j,i}$ and the corresponding $C_{j,i}$ are shown for transitions between
two and three and three and four electrons in the dot, respectively. The
truncation of the Hilbert space to $M\le13$ leads to an absolute error of
$10^{-3}$ (table \ref{table23}) or $10^{-2}$ (table \ref{table34}). In a few
cases we checked the improvement achieved by using $M\!=\!14$ single
electron levels.

Some transitions are almost completely suppressed in the case of the
$M_{j,i}$ compared to the $C_{j,i}$. The two reasons can be seen from
 \begin{equation}
\left|\sum_{n,\sigma} \langle \Psi_{j}| c_{n,\sigma}^{\dagger}
|\Psi_{i}\rangle\right|^{2} = \left|\sum_{\nu_i,\nu_j} \mbox{b}_{\nu_i}
\mbox{b}_{\nu_j} \sum_{n,\sigma} \langle \nu_{j}|
c_{n,\sigma}^{\dagger} |\nu_{i}\rangle\right|^{2} \;.
 \end{equation}
Firstly it may be impossible to create $\langle \nu_{j}|$ by adding one
electron to $|\nu_{i}\rangle$ (spatial selection rule) for the largest
$|\mbox{b}_{\nu_i}\mbox{b}_{\nu_j}|$. Secondly, the various contributions to
the summation over $\nu_i$ and $\nu_j$ may cancel due to different signs of
$\mbox{b}_{\nu_i}\mbox{b}_{\nu_j}$. This latter cancellation seems to have
been neglected in \cite{daniela95}. As an example, table \ref{coef34} shows
the four main contributions for the $|\Psi_0^{1/2}\rangle$ for $N\!=\!3$ and
$|\Psi_3^{0}\rangle$ for $N\!=\!4$ together with their corresponding
electronic configurations (see table \ref{spectrum} for the notation). They
are needed to calculate $M_{3,0}$ in table \ref{table34}. Of the sixteen
possible transitions shown in table \ref{coef34}, only two (marked with $^*$
and $^{\ddagger}$) fulfill the spatial selection rule. In addition their
contribution to $M_{3,0}$ cancel each other due to opposite signs
($M_{3,0}\!\approx\!(-0.35\!\times\!0.29 +
0.33\!\times\!0.39)^2\!=\!0.0006$). This explains the smallness of the
corresponding transition probability. An other example is the transition
between the first excited state for $N\!=\!3$ and the groundstate for
$N\!=\!2$, $\!\langle\Psi_1^{S'=1/2}| c^{\dagger} |\Psi_0^{S=0}\rangle$,
table \ref{table23}. In this case, we were able to follow the evolution of
the many--electron states down to zero interaction, ending at Slater
determinants $|\nu_i\rangle$ and $|\nu_j\rangle$ between which the spatial
selection rule forbids transitions.

\begin{figure}[h]
\begin{center}
\unitlength1in
\begin{picture}(3.4,1.8)
\put(-0.23,-0.35){\includegraphics{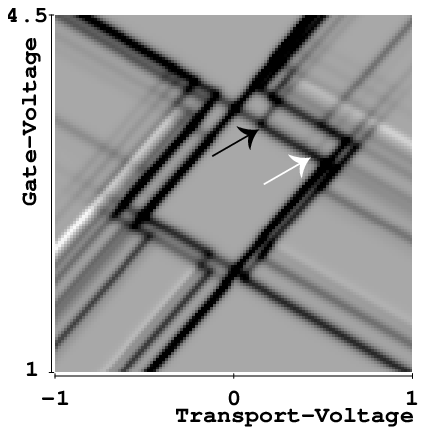}}
\put(1.45,-0.35){\includegraphics{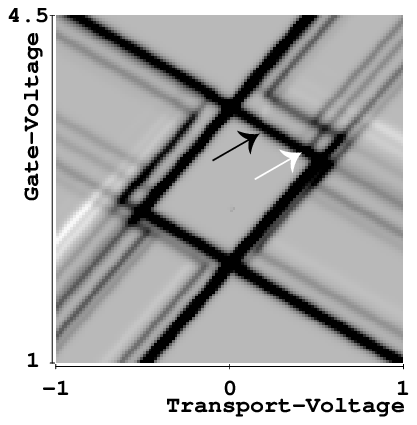}}
\end{picture}
{\caption[greys]{\label{greys} Differential conductance versus transport
and gate voltages (in units of $e/a^*_{\rm B}$) in linear grey scale (dark:
positive; bright: negative). The electron number inside the diamond shaped
Coulomb blockade region is $N\!=\!3$. Left: transition probabilities
proportional to $C_{j,i}$. Right: transition probabilities proportional to
$M_{j,i}$. The arrows are explained in the text.}}
 \end{center}
 \end{figure}

The effect on {\it non--linear transport properties} is demonstrated in
figure \ref{greys}. It shows grey scale plots of the differential
conductance versus the gate voltage $V_{\rm G}$ and the transport voltage.
In figure \ref{greys}(left) the $\Gamma_{j,i}^{L/R}$ were assumed to be
proportional to $C_{j,i}$ \cite{dietmar94} while the calculated $M_{j,i}$
(eq. \ref{matrat}) were used in figure \ref{greys}(right). Grey areas
correspond to regions of zero differential conductance. Black and white
lines are related to positive and negative differential conductances,
respectively (spin blockade \cite{dietmar94,dietmar95}). They reflect
excited many electron states that become available for transport when gate
and/or bias voltages are increased. On average, the number of lines is
reduced in figure \ref{greys}(right) as compared to figure
\ref{greys}(left) (eg. black arrows). This reflects the suppression of
transition matrix elements by the spatial selection rule discussed above. In
some regions, however, the differential conductance is even enhanced (cf.
white arrows in figure \ref{greys}). This is caused by considerable
upheavals in the stationary occupation probabilities $P_j$ obtained from the
rate equation when the full matrix elements are considered in (\ref{matrat}).

\begin{figure}[h]
\begin{center}
\unitlength1cm
\begin{picture}(8,4.5)
\put(0.0,4.3){$I/(e\bar{t})$}
\put(6.15,0.18){$eV_{\rm G}/E_{\rm H}$}
\put(-2.3,-1.2){\includegraphics{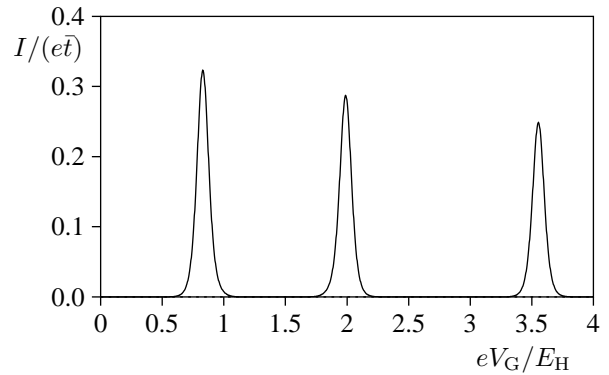}}
\end{picture}
{\caption[lincon]{\label{lincon} Current versus gate voltage for small
transport voltage ($V<\Delta$) using eq.~(\ref{matrat}). Temperature is
$10mK$. The first peak corresponds to oscillations between $N\!=\!1$ and 2
electrons.}}
 \end{center}
 \end{figure}

The conductance peaks at low transport voltages shows different peak heights
as presented in figure \ref{lincon}. This is directly related to the spatial
properties of the many body states. Other works \cite{stone93,falko94}
explain this feature, also observed experimentally \cite{johnson92,weis93},
within the framework of noninteracting electrons picture by semiclassical
chaotic motions. In how far this picture can be generalized to the
correlated electron situation deserves further research. Similar results in
the presence of magnetic fields were shown in \cite{palacios93}.

In summary, we have studied the electron transport through a quantum dot
taking fully into account the correlated eigenstates of the interacting
electrons inside the dot. The spatial selection rule is shown to explain the
suppression of certain transitions between $N$ and ($N\!\pm\!1$) electron
states that would be allowed when taking into account only the spin
selection rules. Despite the obtained tendency towards reduced transition
probabilities $M_{j,i}$ some of the peaks in the differential conductance
are even enhanced as a result of considerable changes in the stationary
occupation probabilities. Furthermore, the correlations between the electrons
induced by the Coulomb interaction lead to characteristic variations in the
heights of the linear conductance peaks. Our results show that non--linear
transport spectroscopy provides in principle valuable information about the
correlated dot states. To extract this information, and the corresponding
physics, however, very careful theoretical modelling is required.

This work was supported by grants of the Deutsche Forschungsgemeinschaft
and the EEC (Contracts No.SCC$^{*}$--CT90--0020 and CHRX--CT930126).

\vspace{-1cm}

\fussy

\begin{table}[h]
\begin{center}
\begin{tabular}{@{\hspace{0.25cm}}c@{\hspace{0.25cm}}|
@{\hspace{0.25cm}}c@{\hspace{0.25cm}}|@{\hspace{0.25cm}}c@{\hspace{0.25cm}}}
\hline
$N=2$ & $N=3$ & $N=4$ \\ \hline
$E_0^0$, $E_1^1$ &
$E_0^{1/2}$, $E_1^{1/2}$, $E_2^{3/2}$ &
$E_0^0,\;E_1^1,\;E_2^1,\;E_3^0,\;E_4^1,\;E_5^2$\\ \hline
\end{tabular}
\end{center}
\caption[table]{\label{spectrum} Sequence of increasing energy eigenvalues
$E_{\alpha}^S$ together with their total spins $S$.}
 \end{table}

 \begin{table}[h]
 \begin{center}\tabcolsep1mm
 \begin{tabular}{rcl||c|c}
\hline
 \hspace{0.5cm}$\langle\Psi_{j(N=3)}^{S\pm1/2}|$ & $c^{\dagger}$ &
$|\Psi_{i(N=2)}^{S} \rangle$\hspace{0.5cm}
&\hspace{0.3cm}$M_{j,i}$\hspace{0.3cm} &
\hspace{0.3cm}$C_{j,i}$\hspace{0.3cm} \\ \hline \hline
$\langle\Psi_0^{1/2}|$ & $c^{\dagger}$& $|\Psi_0^0\rangle$&0.85&\\
$\langle\Psi_1^{1/2}|$ & $c^{\dagger}$& $|\Psi_0^0\rangle$&0.04&
\raisebox{2ex}[-2ex]{1}\\ \hline $\langle\Psi_0^{1/2}|$ & $c^{\dagger}$&
$|\Psi_1^1\rangle$&0.32&\\ $\langle\Psi_1^{1/2}|$ & $c^{\dagger}$&
$|\Psi_1^1\rangle$&0.29& \raisebox{2ex}[-2ex]{1/3}\\ \hline
$\langle\Psi_2^{3/2}|$ & $c^{\dagger}$ & $|\Psi_1^1\rangle$ & 0.43 & $2/3$
\\ \hline
 \end{tabular} \end{center}
 {\caption[table]{\label{table23} Comparison between numerically calculated
matrix elements $M_{j,i}$, eq.(\ref{overmatel}), and corresponding values
obtained by neglecting the spatial part of the wave function, $C_{j,i}$,
eq.~(\ref{clebgor}), for the non--vanishing transition probabilities
$|\Psi_{i(N=2)}^{S} \rangle \rightarrow |\Psi_{j(N=3)}^{S\pm1/2}\rangle$.}}
 \end{table}

\begin{table}[h]
\begin{center}\tabcolsep1mm
\begin{tabular}{rcl||c|c}
\hline
\hspace{0.3cm}$\langle\Psi_{j(N=4)}^{S\pm1/2}|$ & $c^{\dagger}$ &
$|\Psi_{i(N=3)}^{S}\rangle$\hspace*{0.3cm}
&\hspace{0.3cm}$M_{j,i}$\hspace{0.3cm} & \hspace{0.3cm}$C_{j,i}$\hspace{0.3cm}
\\ \hline \hline
$\langle\Psi_0^0|$ & $c^{\dagger}$ &
$[|\Psi_0^{1/2}\rangle;|\Psi_1^{1/2}\rangle]$ & [~0.37 ; 0.15~] & \\
$\langle\Psi_3^0|$ & $c^{\dagger}$ &
$[|\Psi_0^{1/2}\rangle;|\Psi_1^{1/2}\rangle]$ & [~0.01 ; 0.10~] &
\raisebox{2ex}[-2ex]{1/4}
\\ \hline
$\langle\Psi_1^1|$ & $c^{\dagger}$ &
$[|\Psi_0^{1/2}\rangle;|\Psi_1^{1/2}\rangle]$ & [~0.37 ; 0.11~] & \\
$\langle\Psi_2^1|$ & $c^{\dagger}$ &
$[|\Psi_0^{1/2}\rangle;|\Psi_1^{1/2}\rangle]$ & [~0.03 ; 0.49~] & 3/4\\
$\langle\Psi_4^1|$ & $c^{\dagger}$ &
$[|\Psi_0^{1/2}\rangle;|\Psi_1^{1/2}\rangle]$ & [~0.00 ; 0.16~] &
\\ \hline
$\langle\Psi_1^1|$ & $c^{\dagger}$ &
$|\Psi_2^{3/2}\rangle$ & 0.28 & \\
$\langle\Psi_2^1|$ & $c^{\dagger}$ &
$|\Psi_2^{3/2}\rangle$ & 0.23 & 3/8 \\
$\langle\Psi_4^1|$ & $c^{\dagger}$ &
$|\Psi_2^{3/2}\rangle$ & 0.15 &
\\ \hline
$\langle\Psi_5^2|$ & $c^{\dagger}$ & $|\Psi_3^{3/2}\rangle$ & 0.41 & 5/8
\\ \hline
\end{tabular}
\end{center}
\caption[table]{\label{table34} Same as table \ref{table23} for transitions
$|\Psi_{i(N=3)}^S\rangle \rightarrow |\Psi_{j(N=4)}^{S\pm1/2}\rangle$.
Different columns are used for state with same spins $S$ but different
energies (see table \ref{spectrum}).}
 \end{table}

 \newcommand{\uas}{\scriptstyle\uparrow}
 \newcommand{\das}{\scriptstyle\downarrow}
 \newcommand{\spinos}{\thinlines\put(4.5,4.3){$\das$}}
 \newcommand{\spinot}{\thinlines\put(4.5,8.05){$\das$}}
 \newcommand{\spinofo}{\thinlines\put(4.5,13.3){$\das$}}
 \newcommand{\spinofu}{\thinlines\put(4.5,1.3){$\uas$}}
 \newcommand{\spintf}{\thinlines\put(3.5,1.3){$\das$}
                      \put(5.5,1.3){$\uas$}}
 \newcommand{\spints}{\thinlines\put(3.5,4.3){$\das$}
                      \put(5.5,4.3){$\uas$}}
 \newcommand{\spintt}{\thinlines\put(3.5,8.05){$\das$}
                      \put(5.5,8.05){$\uas$}}
 \newcommand{\spintfo}{\thinlines\put(3.5,13.3){$\das$}
                       \put(5.5,13.3){$\uas$}}

 \newsavebox{\qwellts}
 \savebox{\qwellts}{\unitlength1mm \begin{picture}(10,12) \thicklines
   \put(1,0){\line(1,0){8}} \put(1,0){\line(0,1){12}}
   \put(9,0){\line(0,1){12}} \thinlines \put(1,2){\dashbox{0.4}(8,0){}}
   \put(1,5){\dashbox{0.4}(8,0){}} \put(1,8.75){\dashbox{0.4}(8,0){}}
   \end{picture}}
 \newsavebox{\qwellfs}
 \savebox{\qwellfs}{\unitlength1mm \begin{picture}(10,15) \thicklines
   \put(1,0){\line(1,0){8}} \put(1,0){\line(0,1){15}}
   \put(9,0){\line(0,1){15}} \thinlines \put(1,2){\dashbox{0.4}(8,0){}}
   \put(1,5){\dashbox{0.4}(8,0){}} \put(1,8.75){\dashbox{0.4}(8,0){}}
   \put(1.1,14){\dashbox{0.4}(8,0){}} \end{picture}}

\begin{table}[t]
\begin{center}
\begin{tabular}{p{2cm}|*{3}{c|}c}\hline
\parbox{2cm}{\rule[-1.5ex]{0ex}{5ex}\centering
$\mbox{b}_{\nu}^{\{3,E_0^{1/2},-1/2\}}$} & $-0.64$ & $+0.39^*$ & $-0.32$ &
$-0.29^{\ddagger}$ \\ \hline

\parbox{2cm}{\centering Electronic distribution} &
\parbox{1.5cm}{\centering\unitlength1mm
  \begin{picture}(10,12) \put(0,0){\usebox{\qwellts}} \spintf \spinos
  \end{picture}} &
\parbox{1.5cm}{\centering\unitlength1mm
  \begin{picture}(10,12) \put(0,0){\usebox{\qwellts}} \spinos \spintt
  \end{picture}} &
\parbox{1.5cm}{\centering\unitlength1mm
  \begin{picture}(10,12) \put(0,0){\usebox{\qwellts}} \spinofu \spinos \spinot
  \end{picture}} &
\parbox{1.5cm}{\centering\unitlength1mm
  \begin{picture}(10,18) \put(-1,1.5){
    \begin{picture}(10,15) \put(0,0){\usebox{\qwellfs}} \spintf \spinofo
  \end{picture}} \end{picture}} \\ \hline\hline

\parbox{2cm}{\centering Electronic distribution} &
\parbox{1.5cm}{\centering\unitlength1mm
  \begin{picture}(10,12) \put(0,0){\usebox{\qwellts}} \spintf \spintt
  \end{picture}} &
\parbox{1.5cm}{\centering\unitlength1mm
  \begin{picture}(10,18) \put(-1,1.5){
    \begin{picture}(10,15) \put(0,0){\usebox{\qwellfs}} \spintf \spintfo
  \end{picture}} \end{picture}} &
\parbox{1.5cm}{\centering\unitlength1mm
  \begin{picture}(10,12) \put(0,0){\usebox{\qwellts}} \spints \spintt
  \end{picture}} &
\parbox{1.5cm}{\centering\unitlength1mm
  \begin{picture}(10,18) \put(-1,1.5){
    \begin{picture}(10,15) \put(0,0){\usebox{\qwellfs}} \spints \spintfo
  \end{picture}} \end{picture}} \\ \hline

\parbox{2cm}{\rule[-1.5ex]{0ex}{5ex}\centering
$\mbox{b}_{\nu}^{\{4,E_3^0,0\}}$} & $-0.37$ & $+0.35^{\ddagger}$ & $+0.33^*$ &
$-0.25$ \\ \hline
\end{tabular}
\end{center}
{\caption[]{\label{coef34}
The four largest expansion coefficients $\mbox{b}_{\nu}^{\{N,E_i^S,S_z\}}$
of $|\Psi_0^{1/2}\rangle$ and $|\Psi_3^0\rangle$ needed to calculate
the entry $0.01$ in table \ref{table34}. The basis states $|\nu\rangle$ are
illustrated for $N\!=\!3$, 4 according to the occupations of single electron
levels. Only two transitions between these states are possible by creating
or annihilating one electron, marked with $^*$ or with $^{\ddagger}$.}}
 \end{table}

\end{document}